**TITLE**

Direct pairing of homologous DNA double helices may involve the B-to-C form transition

**AUTHORS**

Alexey K. Mazur[1,2] and Eugene Gladyshev[2]

1. CNRS, Université Paris Cité, UPR 9080, Laboratoire de Biochimie Théorique, 13 rue Pierre et Marie Curie, Paris, France
2. Institut Pasteur, Université Paris Cité, Group Fungal Epigenomics, Paris, France

Correspondence:
alexey@ibpc.fr (A.K.M)
eugene.gladyshev@gmail.com (E.G)

**KEYWORDS**

Recombination-independent, homology recognition, C-DNA

**ABSTRACT**

In many organisms, homologous (or repetitive) chromosomal regions can associate or/and undergo concerted epigenetic changes in the absence of DNA breakage and recombination. The direct specific pairing of DNA duplexes with similar nucleotide sequences represents an attractive mechanism for recognizing such regions. Whereas the pairing of B-DNA duplexes may involve a large energy barrier, C-DNA duplexes are expected to pair much more readily. This unique feature of C-DNA is largely due to the fact that its major groove is wide and very shallow, permitting almost perfect initial homologous contacts between two duplexes without clashing. Overall, the conjectured role of C-DNA in recombination-independent pairing should revive the efforts to understand its structure and function in the cell.

**HIGHLIGHTS**

- Homologous (or repetitive) chromosomal regions can engage in pairing or/and undergo concerted epigenetic changes in the apparent absence of DNA breakage and recombination.
- The direct recognition and pairing of homologous double-stranded DNA (dsDNA) molecules represents an attractive underlying mechanism.
- Direct dsDNA-dsDNA pairing during premeiotic and meiotic silencing in filamentous fungi arguably involves segments of C-DNA.
- The C-DNA conformation is particularly suitable for the direct homologous pairing.
- The B-to-C DNA transition may serve as a low-energy pathway during recombination-independent DNA homology search and recognition.

**OUTSTANDING QUESTIONS**

- How can C-DNA be reliably detected and quantified *in vitro* and *in vivo*?
- Which factors could promote the B-to-C DNA transition *in vivo*?
- Can C-DNA exist prior to the pairing?
- Alternatively, does the B-to-C DNA transition only occur concomitantly with the pairing?

**Recombination-independent homologous pairing: its existence and implications**

In many organisms, homologous (or repetitive) chromosomal regions can engage in pairing or/and undergo concerted epigenetic changes in the apparent absence of DNA breakage and recombination [1]. A large number of such homology-dependent phenomena has been described in animals [1,2]. For example, during mammalian development, selected allelic loci are known to pair transiently, presumably to establish appropriate patterns of gene expression [3]. The association of X chromosomes prior to random X-chromosome inactivation in mammalian females provides a paradigmatic example of such a process [4]. Moreover, mammalian genomes contain large amounts of repetitive DNA normally silenced in the form of constitutive heterochromatin [5]. The initiation of such silencing can occur on newly introduced exogenous tandem repeat arrays and does not require RNA interference [6,7], whereas its pathological misregulation has been implicated in several types of cancer [8] and other disease, such as Type I Facioscapulohumeral muscular dystrophy [9]. The mechanistic basis of homology (or repeat) recognition in all these situations remains unknown.

The classical example of recombination-independent pairing was described in *Drosophila melanogaster* and other Diptera insects [10]. In these animals, homologous chromosomes remain associated in the majority of cell types during development and later in the adult life. The paired state is dynamic [11] and plays a critical role in regulating transvection, a phenomenon in which two alleles comprise one expression unit due to their close physical proximity [10]. Recent studies using haplotype-resolved HiC [12,13] and super-resolution microscopy [14] yielded two important insights. First, it became apparent that the degree of pairing fluctuates substantially along the chromosome lengths, where tightly-paired regions are interrupted by loosely-paired regions. Second, within the tightly-paired regions, homologous segments could not be distinguished even at the highest resolution, hinting at a possibility that such pairing was established at distances corresponding to direct DNA-DNA contacts.

The widespread occurrence of recombination-independent homology-directed phenomena contrasts with the limited understanding of their basis for at least two reasons. First, such processes normally involve large and functionally important genomic regions, which are hard to manipulate and analyze experimentally. Second, the only currently accepted general mechanism of homology recognition between double-stranded DNA (dsDNA) is the Watson-Crick base pairing in cross-hybridization of complementary single strands, which occurs normally after strand breaking and base-pair opening by enzymes involved in recombination. Thus, elucidating mechanistic aspects of recombination-independent pairing requires both better model systems and better understanding of the biophysical properties of DNA.

**Meiosis as a dedicated pairing stage**

Homologous pairing is most apparent in meiosis, a specialized cell division that halves the number of chromosomes to generate gametes [15]. Observable meiotic pairing commences in early prophase I, normally after chromosomes begin to compact and individualize [16]. By late prophase I, homologous chromosomes become closely aligned in a configuration known as synapsis.

Canonically, the synapsis of homologous chromosomes requires the break-generating function of Spo11 [17]. This program is documented in mammals and also in some popular model organisms such as *S. cerevisiae* [18]. Alternatively, homologous chromosomes can pair and fully synapse in the absence of Spo11 [18]. Such pairing takes place in *D. melanogaster* and the roundworm *Caenorhabditis elegans* [18]. It is called recombination-independent because it occurs normally in the absence of Spo11-mediated double-strand breaks [16].

Several hypotheses were put forward to explain the phenomenon of recombination-independent pairing. For example, the pairing was proposed to occur directly at the DNA level via G-quadruplexes [19], other four-stranded DNA structures [20,21], or long-range electrostatic interactions [22]. Alternatively, the pairing was suggested to take place at the level of chromosomal domains relying on the indirect contacts mediated by proteins [23], non-coding RNAs [24] as well as large-scale genomic features such as centromeres, telomeres [25,26] and regions of active transcription [27].

The questions regarding the nature of recombination-independent pairing in meiosis proved non-trivial in part because the behavior of meiotic chromosomes is normally studied by tracing the chromosomal axes and the synaptonemal complex [28], which probably start to develop only after the early recombination-independent pairing has peaked.

**MSUD as a sensitive readout of the early recombination-independent homologous pairing in meiosis**
A meiotic process has been described in a number of filamentous fungi in which unpairable allelic sequences trigger RNAi-mediated silencing in prophase I [29–31]. This process was named "meiotic silencing by unpaired DNA", MSUD [30]. Pioneering studies in *Neurospora crassa* defined some basic properties of unpaired DNA capable of inducing MSUD, including the threshold length of ~1 kbp and the absence of a requirement for promoter activity [30]. Collective efforts of several research groups uncovered many genes involved in MSUD [30], yet the mechanism by which unpaired DNA can be detected remained unknown.

Two aspects of the meiotic program in *N. crassa* and related fungi attest to the high efficiency of MSUD and the associated homology-recognition mechanism. First, these organisms have a haploid premeiotic stage, in which parental nuclei undergo the last round of DNA replication and fuse right before the onset of meiosis [28,32]. Second, the early signs of MSUD (*i.e.*, phenotypic loss-of-function of an unpaired gene) can be detected shortly after the fusion of the haploid nuclei [31]. Yet MSUD still operates normally in the absence of Spo11 and Rad51/Dmc1 proteins, when all meiotic recombination has been essentially eliminated [33].

The fact that MSUD is independent from Spo11 and Rad51/Dmc1 contrasts with the fact that in *N. crassa* the synapsis of homologous chromosomes actually requires Spo11 [34,35]. Thus, while *N. crassa* shares this key features of its meiotic program with other "canonical" systems, it also features a transient recombination-independent process, supporting the possibility that the latter can be a part of the canonical meiotic program as well. Indeed, transient Spo11-independent pairing has been reported in mice [23,36] and yeast [37,38]. Whether this transient pairing is requited for the Spo11-dependent pairing and synapsis remains an open question.

**MSUD shares its basis of homology recognition with repeat-induced point mutation (RIP)**
In some filamentous fungi premeiotic nuclei can engage in two closely-related processes known as "repeat-induced point mutation" (RIP) and "methylation induced premeiotically" (MIP) [39,40]. During RIP and MIP, gene-sized repeats of genomic DNA become subject to strong cytosine-to-thymine mutation (RIP) or cytosine methylation (MIP). The capacity of RIP to recognize repeats irrespective of their relative and absolute positions in the genome (*i.e.*, being on the same or two different chromosomes) as well as their origin, sequence composition and coding potential suggests that a general and very efficient homology search is involved [40].

Studies in *N. crassa* have found that RIP, like MSUD, recognizes repeats by a recombination-independent mechanism [41]. They further uncovered the ability of RIP to detect interspersed homology, *i.e.*, pairs of sequences that share periodically occurring short regions of matching base-pairs interleaved with longer stretches of mismatching base-pairs, over the total length of several hundred base-pairs [41]. Remarkably, interspersed homologies containing matching regions of only 4-bp and corresponding to the overall sequence identify of only 36% can still be recognized by RIP [42].

Recent work in *N. crassa* has shown that RIP and MSUD share the ability to detect interspersed homology [33]. For both processes, the optimal periodicity of matching homologous units is equal to 11 bp, however, the periodicity of 22 bp is also effective [33,41]. Because of such strongly discontinuous sequence identity, the annealing of complementary strands does not appear plausible in these situations. At the same time, the involvement of indirect mechanisms (*e.g.,* those relying on sequence-specific proteins or RNAs) also seems unlikely. Taken together, these considerations point towards the existence of a process that matches intact double-stranded DNA molecules directly.

**Quadruplex-based model of the direct homologous dsDNA-dsDNA pairing**
The major-groove edges of the four Watson-Crick (WC) base-pairs have long been known to possess self-complementary shapes and hydrogen bond valences [43]. This curious chemical property was implicated in some early theories of DNA replication, which postulated the pairing between the nascent and the matrix dsDNAs [44]. This same property provided the basis for a model of the infinite four-stranded complex [45] considered as an intermediate step during homologous recombination [21,46,47].

The current model of the direct dsDNA-dsDNA pairing [48] is based on the same principle, and it was proposed to explain the results of genetic experiments on RIP in *N. crassa* [41]. Specifically, it was shown by computations based on quantum mechanics and molecular dynamics that a major-groove contact between two homologous DNA duplexes requires a stack of 3-4 planar quartets formed by identical WC base-pairs [48]. To allow pairing of long dsDNAs, such contacts must be spaced with a periodicity matching that of the double helices [48]. The model also suggested that the pairing should be more efficient if the participating dsDNAs are folded in a right-handed plectoneme [48]. Subsequently, by setting the interval between the individual contacts to the one observed for RIP and MSUD, it was discovered [33] that the intervening dsDNA segments adopted a C-DNA conformation [49–51]. These results have implicated C-DNA in homologous dsDNA-dsDNA pairing.

**C-DNA: a brief overview**
C-DNA is one of the three canonical right-handed double-helical forms discovered in 1950s by molecular modeling based on different experimental data including X-ray diffraction patterns of crystalline DNA fibers [51–53]. Among the three forms, the B-form was recognized as biologically relevant because its diffraction pattern was observed earlier in sperm heads [54]. The A-form was later found in X-ray structures of protein-DNA complexes [55] and also in bacterial spores and virions [56–58]. In contrast, the biological significance of C-DNA remained elusive. Despite the fact that C-DNA is visibly very different from B-DNA (Fig. 1), all its valence and torsion angles do not go beyond the well-populated zones of the B-DNA family. Therefore, C-DNA is considered as a deformed B-DNA or its close relative [51,59].

Two structural properties of C-DNA suggest its role in the direct dsDNA-dsDNA pairing. First, the low helical pitch of approximately 9 bp allows the two C-DNA duplexes coiled in a right-handed plectoneme to have their major grooves facing each other every 22 bp, matching the optimal spacing of homologous sites in genetic experiments on RIP and MSUD [33,41]. Such dense occurrence of pairing sites may be important for the precision of sequence comparisons. Second, perhaps more importantly, the major groove of C-DNA is wide and very shallow, permitting almost perfect initial homologous contacts without clashing (Fig. 1). Thus, it is reasonable to surmise that the B-to-C DNA transition may provide a low energy pathway for the direct homologous dsDNA-dsDNA recognition and pairing. The mechanism and the driving forces of this transition remain to be elucidated.

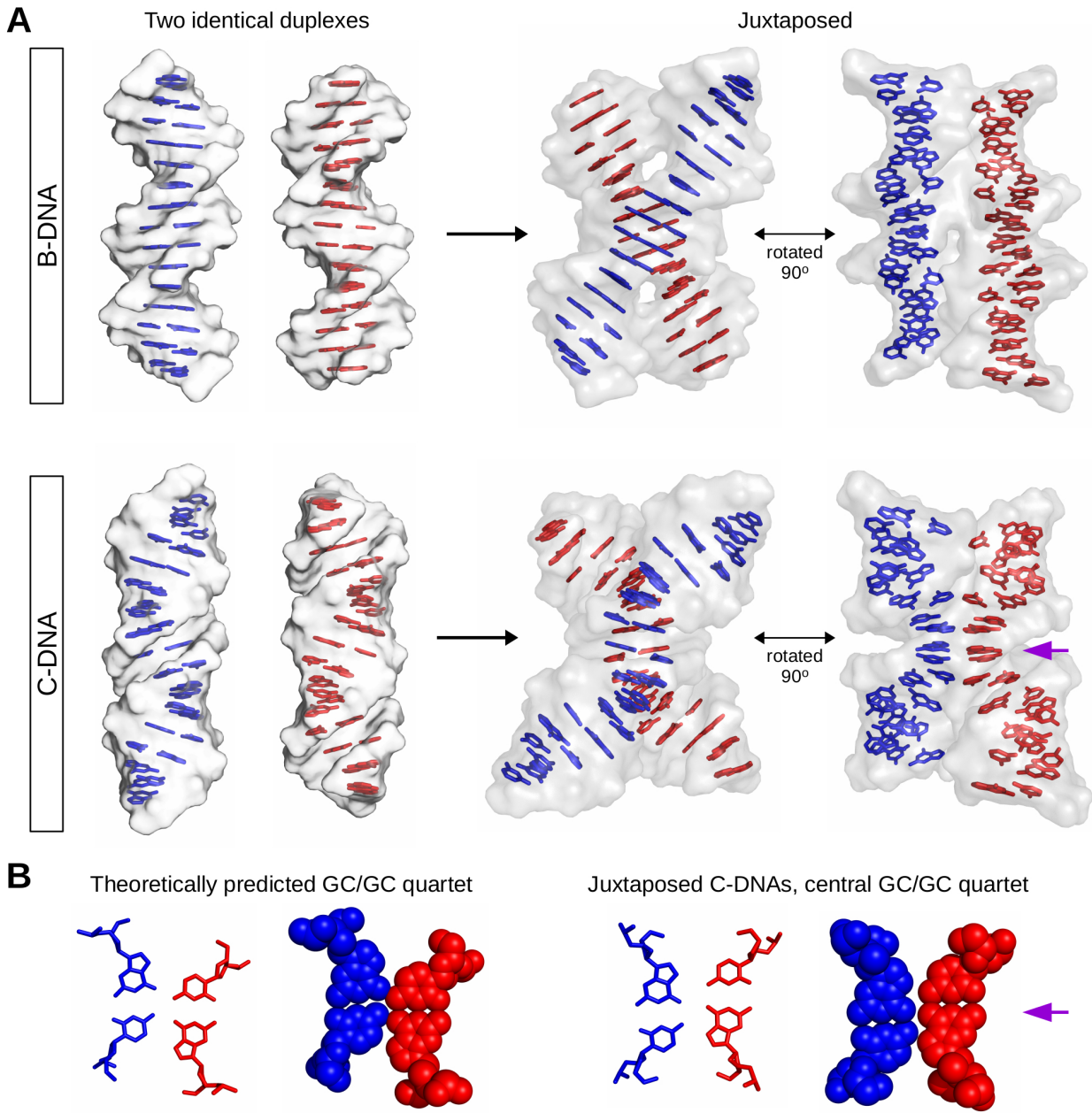

**Figure 1. The possible role of C-DNA in the direct homologous dsDNA-dsDNA pairing.**
**(A)** Canonical B- and C-DNA models are based on X-ray fiber diffraction and solid state NMR data [50,60,61]. Rigid double helices with juxtaposed major grooves were brought together by interactive manual docking with restraints to inter-helix hydrogen bonds of one quartet according to the quadruplex model [48]. The atom-atom clashes were avoided by imposing a minimal inter-phosphate separation of 5.5 Å.
**(B)** The orientation of bases in the central quartet of the paired quadruplex structure produced by all-atom molecular dynamics simulations [48] (left) (left) compared with that between manually docked rigid canonical C-DNA models (right).

Until 1990s, the C-form remained a subject of active investigation by different approaches, both *in vitro* and *in vivo*. For example, it played an important role in the original validation of the Watson-Crick DNA model, based on the fact that the fiber diffraction patterns and the reversible transformations between them could be quantitatively accounted for by conformational transitions between the A, B, and C-forms of the same double helix with unperturbed topology of hydrogen bonding [54]. It was found subsequently that the C-form could be adopted by natural DNA associated with common biological cations [49]. Moreover, only C- and B-forms (but not the A-form) were observed in fibers that were neutralized by charged peptides rather than metal ions [62].

The in-solution B-to-C transitions are typically studied by spectroscopic methods, in particular by circular dichroism (CD) [63–66] The CD method explores the optical activity of chiral substances, thus making it suitable for studying DNA, which is chiral both chemically and topologically. Due to the chiral environment of electrons in stacked bases, the CD spectrum of DNA has a characteristic mid-UV (180-300 nm) band that is very sensitive to transitions between different forms. Because this band does not overlap with the spectra of proteins, the CD method is well applicable *in situ* and *in vivo*. The data interpretation is based on assigning distinct spectral features to known structures [67] because, unfortunately, the CD spectra cannot be reliably calculated from DNA conformations. Based on such correlations, partial B-to-C transitions were reported for DNA in viral particles [63,64], in complexes with lysine-rich proteins or polylysine [68,69] and in chromatin [65,66]. In the case of chromatin, it was further shown that its CD spectrum could be very accurately reproduced by a weighted sum of reference B- and C-form spectra, with the B fraction amounted to 30-50% [70].

These early findings were subsequently challenged by experiments on pure DNA in aqueous solutions. In particular, CD data suggested that gradual non-cooperative B-to-C-DNA transitions could be induced by high concentrations of common salts [65,66,71] or dehydrating cosolvents such as polyethylene glycol and methanol [71,72]. It soon became clear, however, that under similar conditions, concentrated aqueous DNA solutions tended to produce cholesteric liquid crystals, known as "polymer- and salt-induced DNA" (psi- or Ψ-DNA) [73–75]. The Ψ-DNA states yield CD spectra in the same mid-UV range, probably due to intermolecular resonances between many DNA double helices arranged as periodic arrays. Such spectra may adopt different shapes and can have abnormally high amplitudes, regardless of which particular DNA form is present [76,77]. These results questioned the earlier interpretations of CD data for DNA in condensed media as well as in chromatin samples.

In addition, subsequent X-ray experiments showed that DNA in well-hydrated fibers remained in the B-form, regardless of salt concentrations and cosolvents [78]. Similar results revealing the presence of only the B-form were obtained for DNA carrying certain covalent modifications which, according to CD, promoted the B-to-C transition in solution [79]. The last result was not so unambiguous, since it was shown in parallel that, under reduced hydration, this modified sample assumes the C- and not to the A-form as intact DNA. In other words, the structure was indeed pushed towards the C-form, but it was not enough for wet fibers. Nevertheless, the X-ray validation was considered definitive, and it was concluded that the large changes in CD spectra originally attributed to the B-to-C transition actually corresponded to minor conformational changes in B-DNA. In agreement with this idea, the increase of the helical twist of DNA in high-salt aqueous solutions was estimated to be much smaller than that attributed to the B-to-C transition [80].

Once these conclusions became accepted, the interest in C-DNA receded, although the above contradictions could have been interpreted differently. For instance, it became clear that the X-ray diffraction method was not always suitable for verifying in-solution experiments. Specifically, while the Z form was first detected in solution by CD [81,82], and its structure was refined by X-ray analysis in crystals and fibers [83,84], simi-larly to the B-to-C transition, the salt-induced B-to-Z transition could not be reproduced [82]. Overall, transitions to alternative forms were never observed in fibers under high hydration conditions. The probable reason for this is that the environment of DNA in swollen fibers becomes similar to that in solution only when the crystalline order has already been lost.

Furthermore, the helical twist of putative C-DNA in high-salt solutions was measured [80] by an indirect method developed and validated under nearly physiological conditions [85,86]. This method is based on counting the number of ethidium bromide molecules incorporated during the unfolding of natively-supercoiled circular plasmids to planar circles. As the result, it depends on several calibrated parameters and assumptions that cannot be easily transferred to high-salt conditions. For example, as the salt concentration increases, such plasmids undergo a transition to a state known as the dense plectoneme, in which two coiled double helices become strongly and continuously associated along their lengths [87,88]. Unfolding such structures likely involves a series of steps that cannot be all described by a single set of parameters.

The above considerations suggest that the search for the possible biological role of C-DNA could have been interrupted prematurely and warrant further investigations. Specifically, it is puzzling why, given the large amplitude of changes of CD spectra attributed to the B-to-C transition [67], the good agreement between CD and X-ray data observed for A, B, and Z forms breaks down for the C-form. One also needs to keep in mind that while some tests argued against the existence of the C-form in solution, others (for example, the study of covalently modified DNA discussed above) argued otherwise [79]. Our recent results point to the possibility of the B-to-C transition as a way to enable efficient dsDNA-dsDNA pairing [33]. The precise conditions that favor this process need to be clarified and they may include the concerted action of proteins, ions, cosolvents, and supercoiling.

**Generalized implications for meiotic pairing in other organisms**
While MSUD involves a very efficient DNA homology search, it likely represents an extreme instance of a potentially more general process. Indeed, pairing and synapsis of homologs in meiosis can be achieved by using only a small number of dedicated sites [89], and a similar strategy has been conjectured for somatic pairing in *D. melanogaster* [12,13].

As introduced above, recombinational repair of programmed double-strand breaks (DSBs) provides the only known mechanism of homology recognition in meiosis. Meiotic DSBs are typically made by the Top6-like complex that contains Spo11 (corresponding to the Top6A subunit) and the associated Top6B-like subunit [90,91]. While canonical Top6 enzymes can relax supercoiled DNA *in vitro* by cleaving and religating one of the two participating DNA duplexes [92], a reconstituted Spo11/Top6BL complex lacks any catalytic activity [93]. Instead, the properties of the Spo11/Top6BL complex have been studied *in vivo*, where they appear to be controlled by a multitude of factors [94]. Those factors include several processes that imply interactions between potential allelic DSB sites (summarized in [94]). Two types of such interactions have been reported. First, the probability of a DSB at a given site on one chromosome can be affected by the properties of an allelic site on the other homologous chromosome [95,96]. Second, even at very strong hot spots of Spo11/Top6BL activity, only one of the four chromatids is usually broken [97]. These results are consistent with a previously formulated hypothesis that early recombination-independent homologous interactions may guide the break-generating activity of the Spo11/Top6BL complex [38].

The all-atom model of the homologous dsDNA-dsDNA pairing features two DNA duplexes contacting one another every 22 base-pairs [33]. In general, the number of contacts needs to be sufficiently large to differentiate homologous pairing from non-specific associations (*e.g.*, plectonemes and catenanes), but it does not have to be saturating. In this situation, unpaired yet rigidly co-oriented short DNA segments in the vicinity of the paired quadruplexes may represent a specific structure recognized and cleaved by the Spo11/Top6BL complex. In support of this idea, the spacing between closely-positioned concerted cleavage events [98,99], which presumably reports the localization and activity of Spo11/Top6BL, reflects the predicted spacing between the quadruplex contacts [33].

In summary, although the recombination-mediated search currently represents the only known mechanism of DNA homology recognition, the experimental evidence continues to accumulate in support of an alternative, recombination-independent process that is based on the direct pairing of double-stranded DNA molecules. In particular, the energy barrier for the direct dsDNA-dsDNA paring can be decreased if the involved DNAs are present in the C-form rather than the canonical B-form. Further research in this direction is expected to yield new fundamental insights into the mechanism of recombination-independent DNA homology recognition.

**ACKNOWLEDGMENTS**
The work was supported by Agence Nationale de la Recherche (10-LABX-0062, 11-LABX-0011, ANR-19-CE12-0002), Centre National de la Recherche Scientifique (CNRS), and Institut Pasteur.